\documentclass[journal]{IEEEtran}
\usepackage{amsmath}
\usepackage{graphicx}
\usepackage{amssymb}
\usepackage{xcolor}
\usepackage{tabularx}
\usepackage{ragged2e}
\usepackage{booktabs}
\usepackage{stfloats}
\usepackage{array}
\usepackage{makecell}
\usepackage{cite}
\ifCLASSINFOpdf
\else
  
\fi

\begin{document}

\title{From Electromagnetic Physics to Intelligent Optimization: A Deterministic Modeling for Indoor Pinching-Antenna Systems}

\author{Qiushi~Zhao,
        Zihan~Feng,
        Ximing~Xie,~\IEEEmembership{Member,~IEEE},
        and~Hao~Qin,~\IEEEmembership{Member,~IEEE}%
\thanks{Qiushi Zhao is with the School of Optoelectronic Science and
Engineering, University of Electronic Science and Technology of China,
Chengdu 611731, China (202622050112@std.uestc.edu.cn).}%
\thanks{Zihan Feng is with the School of Electrical and Electronic Engineering, Nanyang Technological University,
Singapore 639798, Singapore (FENG0216@e.ntu.edu.sg).}%
\thanks{Ximing Xie is with the Department of Electrical and Computer Engineering, Western University, London, ON N6A 3K7, Canada (xxie269@uwo.ca).}
\thanks{Hao Qin is with the College of Electronics and Information
Engineering, Sichuan University, Chengdu 610065, China (hao.qin@scu.edu.cn).}%
\thanks{Manuscript received XXX; revised XXX.}}

%\markboth{IEEE Transactions on Wireless Communications}%
%{Zhao \MakeLowercase{\textit{et al.}}: From Electromagnetic Physics
%to Intelligent Optimization}

\maketitle

% As a general rule, do not put math, special symbols or citations
% in the abstract or keywords.
\begin{abstract}
Pinching-antenna systems (PASS) enable flexible wireless transmission by activating pinching antennas (PAs) along dielectric waveguides. However, existing PASS analyses commonly rely on tractable analytical or
statistical channel abstractions that do not jointly capture in-waveguide field evolution, sequential PA power extraction, and site-specific indoor multipath propagation. This paper develops a physics-driven deterministic framework for end-to-end modeling of indoor PASS. Specifically, the parabolic wave equation (PWE) resolves the position-dependent electromagnetic field inside the dielectric waveguide, and coupled-mode theory (CMT) maps the local guided field to the radiation power and phase of each PA. Ray tracing (RT) subsequently models the site-specific indoor
electromagnetic propagation from each PA to the users, including LoS, reflected, and diffracted components. To enable efficient repeated evaluations, we combine a reusable PWE-CMT field representation with a physics-assisted neural-network-based RT surrogate that preserves propagation geometry, visibility, and phase information. The proposed framework achieves a complex-channel NMSE of
-19.14 dB while reducing the end-to-end single-link evaluation
time from approximately 2.5 s to 77 ms. Its
application to single- and multi-user deployment demonstrates that the
physical model can be efficiently reused for PA-position and
power-allocation optimization while preserving the underlying
position-dependent propagation physics.
\end{abstract}

% Note that keywords are not normally used for peerreview papers.
\begin{IEEEkeywords}
Pinching-antenna systems, deterministic channel modeling, parabolic wave equation, ray tracing.
\end{IEEEkeywords}

\IEEEpeerreviewmaketitle

\section{Introduction}

Pinching-antenna systems (PASS) have emerged as a flexible-antenna architecture for future wireless communications \cite{10896748,10909665,10912473}. 
Unlike conventional fixed arrays, PASS activates pinching antennas (PAs) at reconfigurable positions along dielectric waveguides to couple guided electromagnetic energy into free space \cite{10945421,11016750}. This positional flexibility, together with a variable number of active PAs, enables channel reshaping over spatial scales much larger than a wavelength. In particular, PAs can be placed near intended users to strengthen line-of-sight (LoS) links and mitigate large-scale propagation loss and blockage \cite{11036558}. Multiple PAs can also be activated along one or more waveguides \cite{11048566,11050939}, without requiring a fully populated antenna array, offering scalable and potentially low-cost spatial transmission. The additional spatial degrees of freedom from PA placement further enable pinching beamforming, where the propagation channel is reconfigured through antenna positioning rather than solely through baseband precoding \cite{11601992,11263923}.

Despite rapid advances in PASS communication theory, most performance analyses still rely on tractable analytical or statistical channel abstractions\cite{10912473,10945421,11016750,11048566,11050939,11036558,11063467,11096622,11098708}. The PA-user link is typically modeled by simplified LoS free-space path loss with spherical-wave phase \cite{10912473,10945421,11016750,11048566,11050939,11063467}, optionally augmented by statistical blockage \cite{11096622} or NLoS multipath \cite{11098708}. Meanwhile, in-waveguide propagation is often reduced to an analytical phase shift, with attenuation neglected\cite{11016750,11048566,11050939,11063467,11098708} or approximated by a prescribed loss coefficient \cite{10912473,11096622}. While enabling closed-form analysis and low-complexity optimization, these models do not explicitly capture site-specific obstacles, reflection, diffraction, and multipath \cite{4578944,8503761}. These channel laws also neglect the longitudinal waveguide-field evolution
and position-dependent sequential power extraction, potentially leading to
suboptimal designs in realistic environments.

Physics-based deterministic channel modeling offers a natural alternative and has been applied to propagation problems with similar physical characteristics. For waveguide-like structures such as railway tunnels, with long longitudinal extent, confined transverse dimensions, and predominantly forward propagation, the parabolic wave equation (PWE) efficiently captures the longitudinal field evolution together with its transverse variation \cite{qin2024comparative,11394627,wu2025intelligentoptimizationwirelessaccess}. In addition, ray tracing (RT) is well established for site-specific indoor propagation, reconstructing LoS, reflected, and diffracted paths from environmental geometry and material properties \cite{8852823,8167305, zhao2024efficient}, particularly at millimeter-wave frequencies where propagation is strongly geometry dependent \cite{11146875}. These methods are complementary for PASS: PWE models the guided in-waveguide field, while RT captures propagation from each PA to users. Their integration enables an end-to-end physical description from waveguide excitation to the received electromagnetic field in PASS.

\subsection{Prior Work}

Early PASS studies focused primarily on establishing the fundamental communication architecture and quantifying the gains offered by pinching antennas \cite{10896748,10909665,10912473,10945421}. Initial experimental demonstrations showed that dielectric particles attached to a waveguide can act as reconfigurable radiating elements\cite{11202577}. 
Building on this concept, theoretical studies investigated single-PA transmission and multi-waveguide configurations, highlighting both the path-loss benefits of flexible PA placement and the additional spatial degrees of freedom enabled by multiple waveguides \cite{10945421}, as well as multi-PA operation along a common waveguide with position-dependent guided phases \cite{10896748,10909665,10912473}. These works established the basic single-waveguide, multi-PA, and multi-waveguide PASS models that form the foundation of the subsequent literature.

A second line of research incorporated PASS into broader communication, estimation, sensing, and wireless-energy applications. The ability of multiple PAs on the same waveguide to radiate a common superimposed signal has motivated studies integrating PASS with non-orthogonal multiple access (NOMA) \cite{11488474}. In parallel, multi-waveguide architectures have been explored to support multi-user transmission and interference management \cite{11165763,zhao2026dynamicantennaplacementmobile}. 
Other studies address practical channel acquisition by developing learning-based estimators that accommodate dynamic PA positioning and varying numbers of active PAs \cite{11364174}. Additionally, PASS was also introduced into integrated sensing and communication (ISAC) systems, where the spatial reconfigurability of PAs was exploited to jointly enhance communication and sensing performance \cite{11551681,11651611}. Other PASS studies further explored areas such as resource allocation \cite{11625963}, localization \cite{11587982}, and wireless power transfer \cite{11106459}. These studies substantially broadened the application scope of PASS, but most retained analytically tractable channel models in order to focus on the associated communication or optimization problem.

More recent studies have shifted attention from application-level extensions 
toward modifications of the PASS hardware and waveguide architecture itself. 
Recent architectural developments include segmented waveguides for shorter
propagation distances and flexible signal aggregation
\cite{11348983,11627933}, center-fed structures for bidirectional
feeding and reduced accumulated loss \cite{11550148}, and
multi-mode waveguides for mode-domain multiplexing
\cite{11593823}. These developments demonstrate that PASS is evolving rapidly from continuous-waveguide architectures toward increasingly sophisticated electromagnetic structures. Nevertheless, accurate and computationally manageable modeling of the physical propagation process remains essential across these different architectures.

\subsection{Motivation and Contributions}

The above developments reveal a gap between increasingly sophisticated PASS designs and the channel models used for their evaluation and optimization. PA relocation simultaneously affects the guided-wave phase, local field and available power inside the waveguide, including the residual power delivered to downstream PAs, as well as the geometry-dependent LoS, reflected, and diffracted paths between each PA and the users. These coupled position-dependent effects are difficult to capture with simplified deterministic or statistical models, potentially leading to unreliable PA placement decisions. PWE and RT provide complementary descriptions of in-waveguide field
evolution and site-specific multipath propagation, respectively
\cite{8271988,11394627,wu2025intelligentoptimizationwirelessaccess,
8852823,8167305,11146875}. Coupled-mode theory (CMT) further models power and phase transfer
between coupled modes \cite{9886072}, while neural surrogates can reduce
the cost of repeated RT evaluations \cite{9774859,11139897}. Motivated by these observations, we develop a physics-driven PASS framework integrating PWE-CMT-RT modeling with physics-assisted surrogate evaluation and communication-oriented optimization, reducing repeated electromagnetic evaluations while retaining the essential propagation physics.

The main contributions of this work are summarized as follows.

\begin{enumerate}

    \item \textit{Physics-based deterministic PASS modeling:}
    We establish a cascaded PWE-CMT-RT framework for deterministic end-to-end modeling of indoor PASS. The PWE resolves the position-dependent guided field and power evolution inside the dielectric waveguide, CMT characterizes PA coupling and the sequential extraction of guided power by multiple PAs, and RT reconstructs the site-specific LoS, reflected, and diffracted propagation paths from each PA to the users. The complex contributions from different propagation paths and PAs are coherently combined, establishing a direct mapping from the waveguide excitation and PA configuration to the user received power and spatial radio map.

    \item \textit{Hybrid physics-neural electromagnetic acceleration:}
We develop reusable PWE-CMT field representations and a physics-assisted
path-level RT surrogate to reduce repeated electromagnetic evaluations.
The surrogate incorporates analytical propagation topology, geometric
visibility, path length, and carrier-phase reconstruction, while reference
RT is retained for final candidate refinement.

    \item \textit{Physics-driven single-user PA optimization: }For the single-user optimization problem, we apply the deterministic PWE–CMT–RT framework to PA placement and employ Bayesian optimization to identify favorable PA locations with a limited number of electromagnetic evaluations.

\item \textit{Physics-driven multi-user joint optimization:} For the multi-user optimization problem, we formulate a QoS-constrained joint PA-position and power-allocation problem. A surrogate-assisted supervised coarse-to-fine framework is developed to identify promising PA configurations, followed by reference-RT refinement of the shortlisted candidates.

\end{enumerate}

The remainder of this paper is organized as follows. 
Section~II establishes the deterministic PWE-CMT-RT framework for end-to-end 
PASS channel modeling, while Section~III develops the corresponding 
physics-assisted acceleration and single- and multi-user optimization methods. 
Section~IV evaluates the modeling accuracy, computational efficiency, and 
optimization performance through numerical simulations. Finally, Section~V 
concludes the paper.

\section{Physics-Based Deterministic System Modeling for PASS}
\label{sec:physics_modeling}

\begin{figure*}[t]
    \centering
    \includegraphics[width=0.85\textwidth]{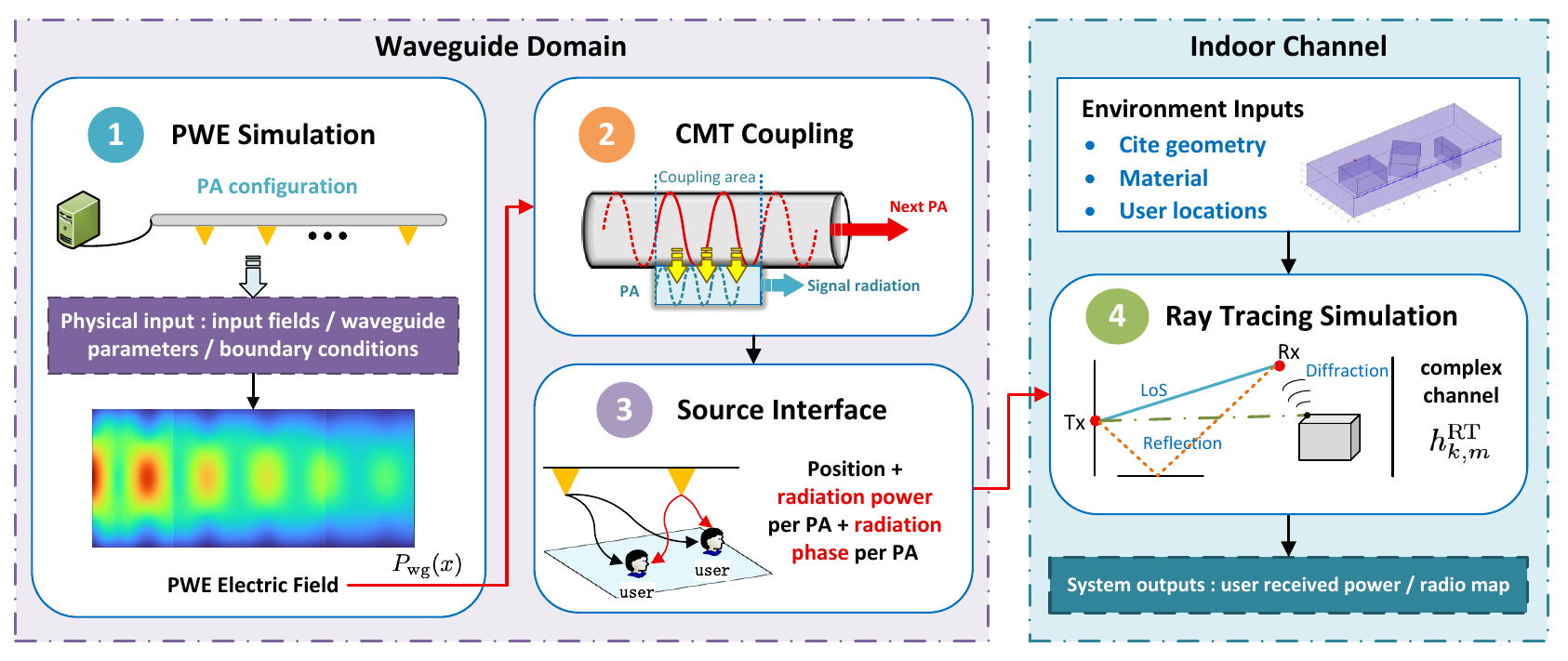}
    \caption{Cascaded physics-driven modeling framework for PASS, including coupled-mode-theory-based PA radiation modeling, PWE-based in-waveguide field simulation, and ray-tracing-based indoor propagation modeling.}
    \label{fig:pass_framework}
\end{figure*}

This section develops a deterministic physics-driven framework that captures the cascaded propagation from the in-waveguide field evolution to the user-side received signal in PASS. Specifically, CMT and PWE are integrated to determine the radiation power and phase of each PA, while RT characterizes indoor multipath propagation and coherently combines the received-field contributions from multiple PAs. The overall modeling framework is illustrated in Fig.~\ref{fig:pass_framework}.

\subsection{Coupled-Mode-Theory-Based Radiation Model}

In PASS, an attached dielectric PA couples a portion of the guided
electromagnetic field from the waveguide and radiates it into free space. We model this power transfer as coupling between the waveguide and PA modes using CMT.

We consider a dielectric waveguide deployed along the $x$-axis. The transverse coordinates are denoted by $y$ and $z$. Let $n_g$ and $n_p$ denote the effective refractive indices of the waveguide and the PA, respectively. The corresponding propagation constants are
\begin{equation}
    \beta_g = k_0 n_g, \qquad \beta_p = k_0 n_p,
\end{equation}
where $k_0=2\pi/\lambda$ is the free-space wavenumber and $\lambda$ is the free-space wavelength. The phase mismatch between the two coupled structures is defined as
\begin{equation}
    \Delta\beta=\beta_g-\beta_p.
\end{equation}
When $\Delta\beta=0$, the fields in the dielectric waveguide and the PA accumulate the same phase over the same propagation distance. Their relative phase therefore remains unchanged throughout the coupling region, allowing the power transferred at different positions to add constructively. When $\Delta\beta\neq0$, a varying phase difference develops and weakens the overall power transfer. Hence, we consider the phase-matched case, i.e., $\beta_g=\beta_p$, which provides the highest coupling efficiency for a given coupling strength and interaction length.

Let $A(\ell)$ and $B(\ell)$ denote the normalized complex modal amplitudes in the waveguide and the PA, respectively, where $\ell\in[0,L]$ is the local coordinate along the PA coupling region and $L$ is the effective coupling length. Under the phase-matched condition, the coupled-mode equations are written as
\begin{equation}
    \frac{dA(\ell)}{d\ell} = -j\kappa B(\ell),
    \label{eq:cmt_A}
\end{equation}
\begin{equation}
    \frac{dB(\ell)}{d\ell} = -j\kappa A(\ell),
    \label{eq:cmt_B}
\end{equation}
where $\kappa$ is the mode-coupling coefficient determined by the
transverse-field overlap. With excitation initially confined to the
waveguide, the initial conditions are
\begin{equation}
    A(0)=1, \qquad B(0)=0.
\end{equation}
Solving \eqref{eq:cmt_A} and \eqref{eq:cmt_B} yields
\begin{equation}
    A(\ell)=\cos(\kappa \ell),
    \label{eq:A_solution}
\end{equation}
\begin{equation}
    B(\ell)=-j\sin(\kappa \ell).
    \label{eq:B_solution}
\end{equation}
At the end of the coupling region, i.e., $\ell=L$, we define the PA field coupling coefficient as
\begin{equation}
    \delta \triangleq \sin(\kappa L).
    \label{eq:delta_single}
\end{equation}
Accordingly, the coupled field amplitude in the PA is scaled by $-j\delta$, while the remaining field amplitude in the main waveguide is scaled by $\sqrt{1-\delta^2}$.

Therefore, for a single PA placed at the longitudinal coordinate $x_p$, the CMT-based local power exchange is expressed as
\begin{equation}
    P_{\mathrm{rad}}
    =
    \delta^2 P_{\mathrm{wg}}^{-}(x_p),
    \label{eq:single_pa_prad}
\end{equation}
\begin{equation}
    P_{\mathrm{wg}}^{+}(x_p)
    =
    (1-\delta^2)P_{\mathrm{wg}}^{-}(x_p),
    \label{eq:single_pa_remain}
\end{equation}
where $P_{\mathrm{rad}}$ denotes the power coupled from the dielectric waveguide and radiated by the PA, while $P_{\mathrm{wg}}^{-}(x_p)$ and $P_{\mathrm{wg}}^{+}(x_p)$ denote the guided powers immediately before and after the PA coupling region, respectively. Hence, $\delta^2$ represents the fraction of the local guided power extracted and radiated by the PA, while $1-\delta^2$ represents the fraction that remains in the waveguide and continues to propagate.

For the multiple-PA case, the sequential nature of the coupling process must be explicitly considered. Suppose that $M$ PAs are attached to the same waveguide in ascending order of their longitudinal coordinates, i.e.,
\begin{equation}
    x_1 < x_2 < \cdots < x_M .
\end{equation}
The position of the $m$-th PA is denoted by
\begin{equation}
    \mathbf{p}_m=[x_m,y_g,z_g]^T,
\end{equation}
where $x_m$ is the longitudinal coordinate of the $m$-th PA, and $y_g$ and $z_g$ are the transverse coordinates of the waveguide. Let $L_m$ be the effective coupling length of the $m$-th PA. Under the phase-matched condition, CMT shows that the fraction of power transferred over the coupling length $L_m$ is $\sin^2(\kappa L_m)$. Accordingly, the amplitude coupling coefficient of the $m$-th PA is defined as
\begin{equation}
    \delta_m \triangleq \sin(\kappa L_m),
    \label{eq:delta_m}
\end{equation}
such that $\delta_m^2$ represents the fraction of the local guided power extracted by the $m$-th PA.

For each $m\in\{1,\ldots,M\}$, the radiation power of the $m$-th PA is given by
\begin{equation}
    P_{\mathrm{rad},m}
    =
    \delta_m^2 P_{\mathrm{wg}}^{-}(x_m),
    \label{eq:multi_pa_prad_local}
\end{equation}
where $P_{\mathrm{rad},m}$ is the radiation power of the $m$-th PA, and $P_{\mathrm{wg}}^{-}(x_m)$ is the guided power immediately before its coupling region. The guided power immediately after the coupling region is
\begin{equation}
    P_{\mathrm{wg}}^{+}(x_m)
    =
    (1-\delta_m^2)P_{\mathrm{wg}}^{-}(x_m),
    \label{eq:multi_pa_remaining_power}
\end{equation}
where $P_{\mathrm{wg}}^{+}(x_m)$ denotes the remaining guided power immediately after the $m$-th PA.

For each $m\in\{2,\ldots,M\}$, the power-transmission factor from $x_{m-1}^{+}$ to $x_m^{-}$ is defined as
\begin{equation}
    \eta_{m-1,m}
    =
    \frac{P_{\mathrm{wg}}^{-}(x_m)}
         {P_{\mathrm{wg}}^{+}(x_{m-1})},
    \label{eq:waveguide_transmission_factor}
\end{equation}
where $\eta_{m-1,m}$ is evaluated from the guided powers computed by the PWE. The incident guided powers at two consecutive PAs are therefore related by
\begin{equation}
\begin{aligned}
    P_{\mathrm{wg}}^{-}(x_m)=
    \eta_{m-1,m}(1-\delta_{m-1}^2) P_{\mathrm{wg}}^{-}(x_{m-1}).
\end{aligned}
\label{eq:multi_pa_power_recursion}
\end{equation}

By recursively applying \eqref{eq:multi_pa_power_recursion}, the radiation power of the $m$-th PA, for each $m\in\{2,\ldots,M\}$, can be expressed in the cascaded form
\begin{equation}
\begin{aligned}
    P_{\mathrm{rad},m}
    &=
    \delta_m^2 P_{\mathrm{wg}}^{-}(x_1)
    \prod_{i=1}^{m-1}
    \Big[
        \eta_{i,i+1}(1-\delta_i^2)
    \Big].
\end{aligned}
\label{eq:multi_pa_cascaded_power}
\end{equation}
For $m=1$,
$P_{\mathrm{rad},1}=\delta_1^2P_{\mathrm{wg}}^{-}(x_1)$.
Equation~\eqref{eq:multi_pa_cascaded_power} shows that the radiation power of each PA is jointly determined by its own coupling coefficient, the power extracted by all preceding PAs, and the PWE-computed propagation between adjacent PA positions.

\subsection{PWE-Based Waveguide Propagation Modeling}

\begin{figure}[h]
    \centering
    \includegraphics[width=0.45\textwidth]{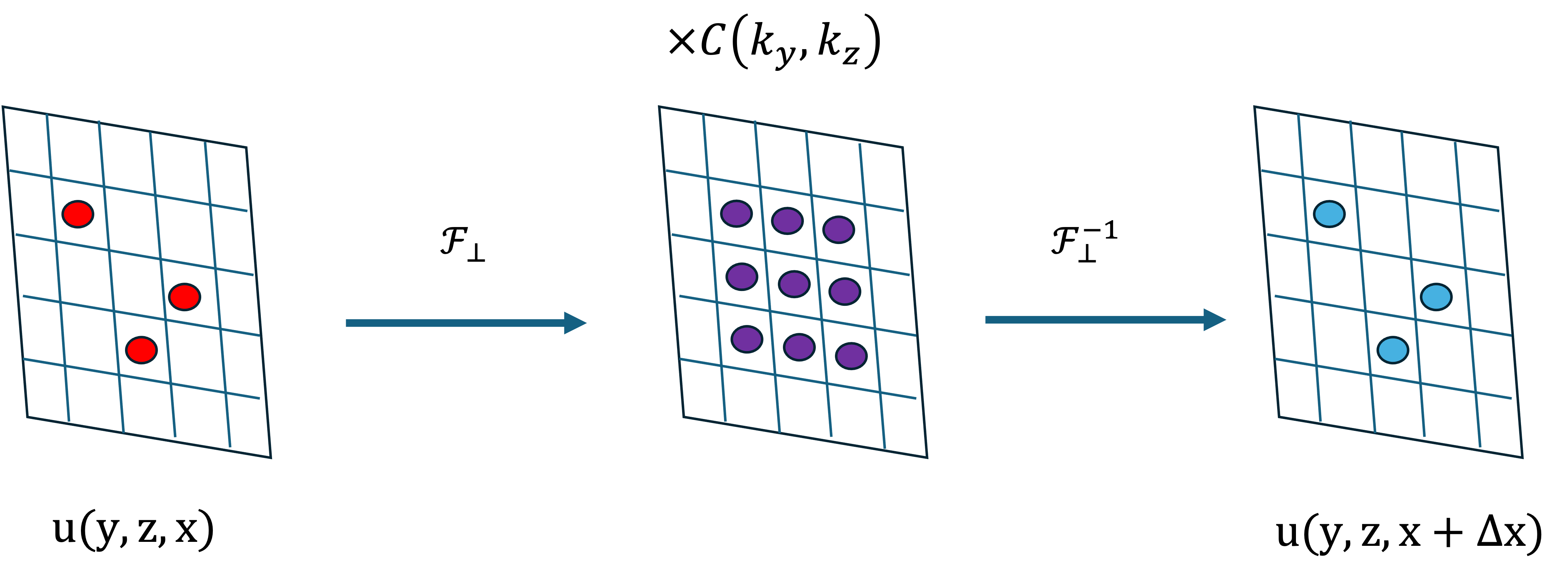}
    \caption{Illustration of the PWE-based waveguide propagation model.}
    \label{fig:principle of PWE}
\end{figure}
Given the local guided power, CMT determines the power extracted by each
PA. We employ the parabolic wave equation to efficiently compute
the predominantly forward-propagating field along the dielectric
waveguide.

The derivation starts from the source-free scalar Helmholtz equation
\begin{equation}
    \left(
    \frac{\partial^2}{\partial x^2}
    +
    \frac{\partial^2}{\partial y^2}
    +
    \frac{\partial^2}{\partial z^2}
    +
    k_0^2 n^2(y,z)
    \right)
    \psi(x,y,z)
    =
    0,
    \label{eq:helmholtz}
\end{equation}
where $\psi(x,y,z)$ denotes the scalar field component of interest, and $n(y,z)$ is the transverse refractive-index profile of the waveguide. The waveguide is assumed to guide the wave predominantly along the $x$-axis. By separating the rapidly varying longitudinal phase, the field is expressed as
\begin{equation}
    \psi(x,y,z)
    =
    u(x,y,z)e^{-jk_{\mathrm{ref}}x},
    \label{eq:pwe_ansatz}
\end{equation}
where $u(x,y,z)$ denotes the slowly varying complex field amplitude after removing the rapidly oscillating reference phase term $e^{-jk_{\mathrm{ref}}x}$, and $k_{\mathrm{ref}}=k_0 n_{\mathrm{ref}}$ is the reference propagation constant. Substituting \eqref{eq:pwe_ansatz} into \eqref{eq:helmholtz} gives
\begin{equation}
    \frac{\partial^2 u}{\partial x^2}
    -
    2jk_{\mathrm{ref}}\frac{\partial u}{\partial x}
    +
    \nabla_{\perp}^{2}u
    +
    \left(k_0^2 n^2(y,z)-k_{\mathrm{ref}}^2\right)u
    =
    0,
\end{equation}
where
$\nabla_{\perp}^{2}
=\partial^2/\partial y^2+\partial^2/\partial z^2$
is the transverse Laplacian. Under the paraxial approximation, the field amplitude $u(x,y,z)$ varies slowly along the propagation direction such that
\begin{equation}
    \left|
    \frac{\partial^2 u}{\partial x^2}
    \right|
    \ll
    k_{\mathrm{ref}}
    \left|
    \frac{\partial u}{\partial x}
    \right|.
\end{equation}
The standard PWE formulation is then obtained as
\begin{equation}
    \frac{\partial u}{\partial x}
    =
    \frac{1}{2jk_{\mathrm{ref}}}
    \left[
    \nabla_{\perp}^{2}
    +
    k_0^2 n^2(y,z)
    -
    k_{\mathrm{ref}}^2
    \right]u.
    \label{eq:pwe_general}
\end{equation}
For a locally homogeneous reference medium with $n(y,z)\approx n_{\mathrm{ref}}$, \eqref{eq:pwe_general} reduces to
\begin{equation}
    \frac{\partial u}{\partial x}
    =
    \frac{1}{2jk_{\mathrm{ref}}}
    \nabla_{\perp}^{2}u.
    \label{eq:pwe_simplified}
\end{equation}

For numerical implementation, the slowly varying field amplitude is advanced along the waveguide by a marching scheme. In the split-step Fourier implementation, the field at $x+\Delta x$ is updated from the field at $x$ as
\begin{equation}
    U(y,z,x+\Delta x)
    =
    \mathcal{F}_{\perp}^{-1}
    \left\{
    C(k_y,k_z)
    \mathcal{F}_{\perp}
    \left[
    U(y,z,x)
    \right]
    \right\},
    \label{eq:ssft}
\end{equation}
where $\mathcal{F}_{\perp}$ and $\mathcal{F}_{\perp}^{-1}$ denote the two-dimensional Fourier transform and inverse Fourier transform over the transverse plane, respectively. This implementation scheme is illustrated in Fig. \ref{fig:principle of PWE}. The spectral-domain propagator is
\begin{equation}
    C(k_y,k_z)
    =
    \exp
    \left[
    -j
    \frac{k_y^2+k_z^2}{2k_{\mathrm{ref}}}
    \Delta x
    \right].
    \label{eq:ssft_operator}
\end{equation}

Equation~\eqref{eq:ssft} is solved as a longitudinal initial-value
problem using the transverse input field. At each marching step, an
impedance boundary condition accounts for field confinement and finite
boundary loss at the computational-window boundary.

The output of the PWE solver is the absolute complex electric-field distribution inside the waveguide, denoted by $E_{\mathrm{wg}}(y,z,x)$ in $\mathrm{V/m}$. This complex field distribution retains both the magnitude and phase accumulated during in-waveguide propagation. The guided power over the transverse cross section $S_{\perp}$ at coordinate $x$ is evaluated using the simplified time-averaged Poynting-flux approximation
\begin{equation}
    P_{\mathrm{wg}}(x)
    \approx
    \sum_{(y,z)\in S_{\perp}}
    \frac{
        \left|E_{\mathrm{wg}}(y,z,x)\right|^2
    }{
        2Z_{\mathrm{eff}}
    }
    \Delta y \Delta z,
    \label{eq:pwe_power}
\end{equation}
where $Z_{\mathrm{eff}}$ is the effective wave impedance of the guided mode, and $\Delta y\Delta z$ is the transverse cell area. When the modal impedance is unavailable, we use
$Z_{\mathrm{eff}}\approx Z_0/n_{\mathrm{eff}}$, where
$Z_0\simeq377~\Omega$ and $n_{\mathrm{eff}}$ is the guided-mode
effective refractive index.

In addition to the guided power, the full complex electric field obtained from the PWE is used to determine the excitation phase of each PA. Because $E_{\mathrm{wg}}(y,z,x)$ includes the restored longitudinal reference carrier, its phase contains the complete phase accumulated during in-waveguide propagation. Let $E_g(y,z)$ denote the transverse field profile of the guided mode, defined using a fixed phase reference. The complex guided-mode coefficient immediately before the coupling region of the $m$-th PA is obtained through modal projection as
\begin{equation}
\begin{aligned}
    a_{\mathrm{wg},m}^{-}
    &=
    \frac{
        \displaystyle
        \sum_{(y,z)\in S_{\perp}}
        E_{\mathrm{wg}}(y,z,x_m^{-})
        E_g^{*}(y,z)
        \Delta y\Delta z
    }{
        \displaystyle
        \sum_{(y,z)\in S_{\perp}}
        |E_g(y,z)|^2
        \Delta y\Delta z
    } .
\end{aligned}
\label{eq:pwe_modal_amplitude}
\end{equation}
Here, $(\cdot)^*$ denotes complex conjugation. The phase of the guided mode incident on the $m$-th PA is therefore
\begin{equation}
    \theta_{\mathrm{wg},m}
    =
    \arg\left(
        a_{\mathrm{wg},m}^{-}
    \right).
    \label{eq:pwe_guided_phase}
\end{equation}

At $x_m^{-}$, \eqref{eq:pwe_power} gives the local guided power,
from which $P_{\mathrm{rad},m}$ follows via
\eqref{eq:multi_pa_prad_local}. Since CMT introduces the complex
coupling factor $-j\delta_m$, the equivalent PA radiation phase is
\begin{equation}
    \theta_{\mathrm{rad},m}
    =
    \arg\left(
        -j\delta_m
        a_{\mathrm{wg},m}^{-}
    \right).
    \label{eq:pa_radiation_phase}
\end{equation}
For a real and positive $\delta_m$, \eqref{eq:pa_radiation_phase} reduces to
\begin{equation}
    \theta_{\mathrm{rad},m}
    =
    \theta_{\mathrm{wg},m}
    -
    \frac{\pi}{2}
    \quad
    (\mathrm{mod}\ 2\pi).
    \label{eq:pa_radiation_phase_reduced}
\end{equation}

The PWE-CMT model consequently provides the position $\mathbf{p}_m$, radiation power $P_{\mathrm{rad},m}$, and radiation phase $\theta_{\mathrm{rad},m}$ of each PA. These quantities define the corresponding equivalent source in the RT model as
\begin{equation}
    \mathcal{S}_m
    \triangleq
    \left\{
        \mathbf{p}_m,
        P_{\mathrm{rad},m},
        \theta_{\mathrm{rad},m}
    \right\},
    \label{eq:pa_source_tuple}
\end{equation}
where $\mathcal{S}_m$ denotes the source-parameter set of the $m$-th PA. Equivalently, the complex source excitation is
\begin{equation}
    q_m
    \triangleq
    \sqrt{P_{\mathrm{rad},m}}\,
    \exp\left(
        j\theta_{\mathrm{rad},m}
    \right).
    \label{eq:pa_complex_excitation}
\end{equation}

\subsection{Ray-Tracing-Based Indoor Channel Modeling}

The PWE-CMT outputs configure each PA as an equivalent RT source
$\mathcal{S}_m$ in \eqref{eq:pa_source_tuple}. Given the site-specific geometry and material properties, the shooting-and-bouncing-rays method identifies the valid paths between each PA and user.

Consider the channel from the $m$-th PA to the $k$-th user located at $\mathbf{u}_k$. Let $\mathcal{P}_{k,m}$ denote the set of valid propagation paths identified by the RT solver. Depending on the configured propagation mechanisms, these paths may include line-of-sight, reflected, and diffracted components. For the $n$-th path in $\mathcal{P}_{k,m}$, the RT solver returns the path loss $\mathrm{PL}_{k,m,n}^{\mathrm{RT}}$ in dB and the propagation phase shift $\phi_{k,m,n}^{\mathrm{RT}}$ in radians. Under the $e^{j\omega t}$ phasor convention, the corresponding complex path gain is
\begin{equation}
\begin{aligned}
    \alpha_{k,m,n}^{\mathrm{RT}}
    =
    10^{-\mathrm{PL}_{k,m,n}^{\mathrm{RT}}/20}
    \exp\left(
        -j\phi_{k,m,n}^{\mathrm{RT}}
    \right),
\end{aligned}
\label{eq:rt_path_gain}
\end{equation}
where $\alpha_{k,m,n}^{\mathrm{RT}}\in\mathbb{C}$ is the dimensionless complex path gain.

For each PA-user pair, the narrowband complex channel coefficient is obtained by coherently combining all valid paths:
\begin{equation}
    h_{k,m}^{\mathrm{RT}}
    =
    \sum_{n\in\mathcal{P}_{k,m}}
    \alpha_{k,m,n}^{\mathrm{RT}},
    \label{eq:rt_channel}
\end{equation}
where $h_{k,m}^{\mathrm{RT}}\in\mathbb{C}$ denotes the complex propagation coefficient from the $m$-th PA to the $k$-th user. It contains the phase shifts introduced by site-specific multipath propagation, whereas the initial radiation phase of the PA is represented separately by $\theta_{\mathrm{rad},m}$. If no valid propagation path exists, $h_{k,m}^{\mathrm{RT}}$ is set to zero. The complete RT channel for the $K$ users and $M$ PAs can therefore be represented by
\begin{equation}
    \mathbf{H}^{\mathrm{RT}}
    =
    \left[
        h_{k,m}^{\mathrm{RT}}
    \right]_{K\times M},
    \label{eq:rt_channel_matrix}
\end{equation}
where $\mathbf{H}^{\mathrm{RT}}\in\mathbb{C}^{K\times M}$ is the site-specific channel matrix.

Let $s$ satisfy $\mathbb{E}[|s|^2]=1$, and define
$q_m=\sqrt{P_{\mathrm{rad},m}}e^{j\theta_{\mathrm{rad},m}}$. The received complex baseband signal at the $k$-th user is
\begin{equation}
y_k=
\sum_{m=1}^{M}q_m h_{k,m}^{\mathrm{RT}}s+n_k.
\label{eq:rt_received_signal}
\end{equation}
where $y_k\in\mathbb{C}$ is the received signal and $n_k\in\mathbb{C}$ is the additive noise at the $k$-th user. Accordingly, the noise-free received signal power at the $k$-th user is
\begin{equation}
\begin{aligned}
    P_{r,k}^{\mathrm{RT}}
    &=
    \left|
        \sum_{m=1}^{M}
        \sqrt{P_{\mathrm{rad},m}}\,
        \exp\left(
            j\theta_{\mathrm{rad},m}
        \right)
        h_{k,m}^{\mathrm{RT}}
    \right|^2 .
\end{aligned}
\label{eq:rt_received_power}
\end{equation}
Here, $P_{r,k}^{\mathrm{RT}}$ denotes the noise-free received signal power at the $k$-th user, measured in watts.

Equation~\eqref{eq:rt_received_power} coherently combines the PA radiation phases and RT-derived multipath phases, thereby retaining constructive and destructive interference.

\subsection{Integrated PASS Model and Radio Map Generation}
\label{subsec:integrated_model}

The integrated PWE-CMT-RT model maps the in-waveguide complex field
distribution to the spatial received power. Let
$h_m^{\mathrm{RT}}(\mathbf{r})$ denote the site-specific complex propagation
coefficient from the $m$-th PA to a sampling point $\mathbf{r}$. The resulting
PASS radio map is
\begin{equation}
\mathcal{R}(\mathbf{r})
=
\left|
\sum_{m=1}^{M}
q_m h_m^{\mathrm{RT}}(\mathbf{r})
\right|^2,
\label{eq:radio_map}
\end{equation}
where $\mathcal{R}(\mathbf{r})$ is the noise-free received power in watts.
Evaluating \eqref{eq:radio_map} over all sampling points yields a
phase-coherent radio map that captures interference induced by both
in-waveguide phase accumulation and site-specific multipath propagation.

% =========================================================
% Section III
% =========================================================
\section{Physics-Driven Communication Optimization for Indoor PASS}
\label{sec:optimization}

\begin{figure*}[t]
    \centering
    \includegraphics[width=0.75\textwidth]{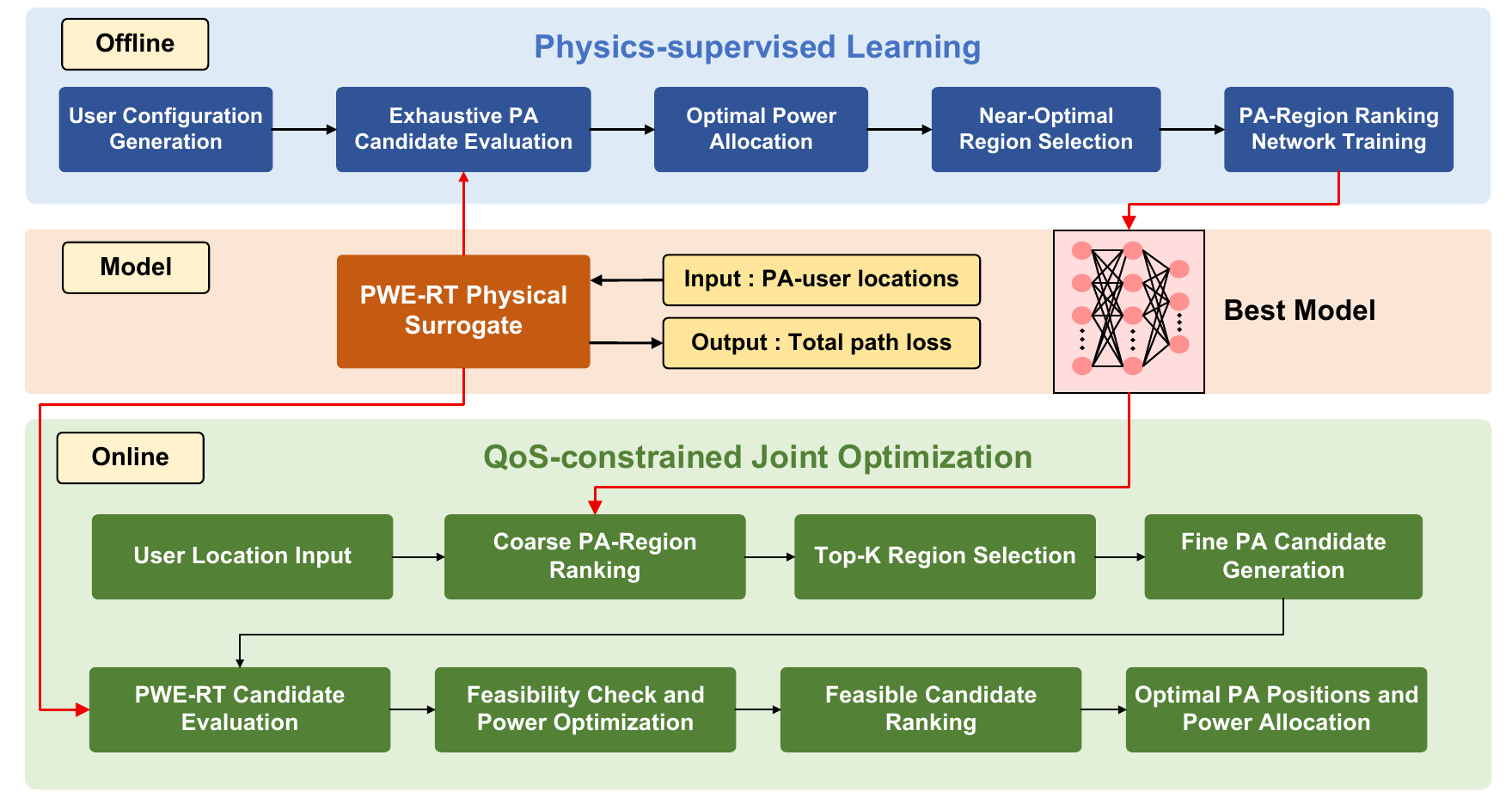}
    \caption{Offline training and online inference workflow of the physics-supervised coarse-to-fine framework for QoS-constrained joint PA-placement and power-allocation optimization.}
    \label{fig:suanfa}
\end{figure*}

Building on the deterministic PWE-CMT-RT framework in
Section~\ref{sec:physics_modeling}, we next investigate its application
to communication-oriented PASS optimization.

\subsection{Single-User PA Position Optimization}
\label{subsec:single_user_optimization}

\subsubsection{Single-User Communication Model}

Consider a single-user indoor PASS consisting of a dielectric
waveguide, one activated PA, and one single-antenna user. The
waveguide is deployed along the $x$-axis, and the longitudinal
position of the PA is denoted by $x_p$. The transverse coordinates
of the PA are fixed by the waveguide installation. Therefore, its
three-dimensional position can be written as
$\mathbf{p}=[x_p,y_g,z_g]^{\mathrm T}$, where $y_g$ and $z_g$
denote the fixed transverse position and installation height of
the waveguide, respectively. The feasible PA position satisfies
$0\leq x_p\leq L_{\mathrm{wg}}$, where $L_{\mathrm{wg}}$ is the
waveguide length.

For a given PA position $x_p$, the deterministic PWE-CMT-RT
framework provides the received power at the fixed user position
$\mathbf{u}_1$, denoted by
$P_{r,1}^{\mathrm{RT}}(x_p,\mathbf{u}_1)$. The corresponding
signal-to-noise ratio (SNR) is
\begin{equation}
    \gamma(x_p)
    =
    \frac{P_r(x_p)}{\sigma^2},
    \label{eq:single_user_snr}
\end{equation}
where $\sigma^2$ denotes the receiver noise power. The achievable
spectral efficiency (SE) is therefore
\begin{equation}
    R(x_p)
    =
    \log_2
    \left(
    1+\frac{P_r(x_p)}{\sigma^2}
    \right)
    \quad \mathrm{bit/s/Hz}.
    \label{eq:single_user_se}
\end{equation}

Therefore, the physics-based deterministic model establishes the
mapping $x_p\longmapsto P_r(x_p)\longmapsto R(x_p)$, which
provides the objective function for the subsequent single-user
PA-position optimization.

\subsubsection{Single-User Optimization Problem}

The single-user PA-position optimization problem is formulated as
\begin{subequations}
\label{prob:single_user_optimization}
\begin{align}
    \mathcal{P}_1:\quad
    \underset{x_p}{\operatorname{maximize}}
    \quad&
    R(x_p)
    \label{prob:single_user_objective}
    \\
    \operatorname{subject\ to}\quad&
    0\leq x_p\leq L_{\mathrm{wg}}.
    \label{prob:single_user_constraint}
\end{align}
\end{subequations}

Since $\log_2(1+x)$ is strictly increasing and the receiver noise
power is independent of $x_p$, maximizing the spectral efficiency
is equivalent to maximizing the received power, i.e.,
$\operatorname{arg\,max}_{x_p}R(x_p)
=\operatorname{arg\,max}_{x_p}P_r(x_p)$. Therefore, the received
power can be directly used as the optimization objective in the
numerical implementation.

Problem~\eqref{prob:single_user_optimization} does not admit an
explicit analytical objective function. Each evaluation of
$R(x_p)$ or $P_r(x_p)$ requires the cascaded PWE-CMT and RT
solvers.
Furthermore, an analytical gradient with respect to $x_p$ is not
available. Consequently, $\mathcal{P}_1$ constitutes a
one-dimensional, deterministic, and computationally expensive
black-box optimization problem.

\subsubsection{Bayesian-Optimization-Based Solution}

Bayesian optimization (BO) is adopted to solve
Problem~\eqref{prob:single_user_optimization} because each
evaluation of the PWE-CMT-RT model is computationally expensive.
Let $f(x_p)\triangleq P_{r,\mathrm{dBm}}(x_p)$ denote the
black-box objective and
$\mathcal{D}_n=\{(x_i,f(x_i))\}_{i=1}^{n}$ the available
observations. A Gaussian process models the objective as
\begin{equation}
    f(x)\mid\mathcal{D}_n
    \sim
    \mathcal{N}\!\left(
        \mu_n(x),
        \sigma_n^2(x)
    \right).
    \label{eq:gp_posterior_single_user}
\end{equation}
Using $f_n^{+}=\max_{1\leq i\leq n}f(x_i)$ and
$z_n(x)=(\mu_n(x)-f_n^{+})/\sigma_n(x)$, the expected improvement
is
\begin{equation}
\begin{aligned}
    \mathrm{EI}_n(x)
    &=
    \left(\mu_n(x)-f_n^{+}\right)
    \Phi_{\mathrm{N}}\!\left(z_n(x)\right)
    +
    \sigma_n(x)
    \varphi_{\mathrm{N}}\!\left(z_n(x)\right).
\end{aligned}
\label{eq:ei_single_user}
\end{equation}
BO iteratively evaluates
\begin{equation}
    x_{n+1}
    =
    \underset{0\leq x\leq L_{\mathrm{wg}}}
    {\operatorname{arg\,max}}\;
    \mathrm{EI}_n(x).
    \label{eq:bo_next_sample}
\end{equation}
After $N_{\mathrm{eval}}$ evaluations, the best evaluated position
is returned as
\begin{equation}
    x_p^{\star}
    =
    \underset{x_i:\,1\leq i\leq N_{\mathrm{eval}}}
    {\operatorname{arg\,max}}\;
    f(x_i).
    \label{eq:bo_final_solution}
\end{equation}

\subsection{Multi-User Joint PA-Position and Power-Allocation Design}
\label{subsec:multi_user_optimization}

\subsubsection{Multi-User PASS Transmission Model}

Consider a PASS comprising a BS, a dielectric waveguide of length
$L_{\mathrm{wg}}$, $M=2$ PAs, and $K$ users. The ordered PA positions
$\mathbf{x}=[x_1,x_2]^{\mathrm T}$ satisfy
$0\leq x_1<x_2\leq L_{\mathrm{wg}}$ and
$x_2-x_1\geq d_{\min}$.

User $k$ is located at
$\mathbf{u}_k=[x_k^{\mathrm u},y_k^{\mathrm u}]^{\mathrm T}$ and height
$h_{\mathrm u}$ within the accessible, obstacle-free region. The complete
user configuration is
$\mathcal U=\{\mathbf{u}_1,\ldots,\mathbf{u}_K\}$.

The users are assumed to share the same time-frequency resource.
Let $s_k$ denote the information symbol intended for user $k$,
satisfying $\mathbb E[|s_k|^2]=1$ and
$\mathbb E[s_i s_j^*]=0$ for $i\neq j$. The normalized composite
baseband signal generated at the BS is
\begin{equation}
    s
    =
    \sum_{k=1}^{K}\sqrt{\mu_k}s_k.
    \label{eq:multi_user_composite_signal}
\end{equation}
Here, $\mu_k$ is the power-allocation coefficient of user $k$.
The coefficients satisfy $\mu_k\geq0$ for all $k$ and
$\sum_{k=1}^{K}\mu_k=1$. The signal injected into the waveguide
is $\sqrt{P_{\max}}s$, where $P_{\max}$ is the total
waveguide-input power. The PWE input field is normalized such that
$P_{\mathrm{wg}}(0)=P_{\max}$.

For a given PA configuration $\mathbf{x}$ and user position
$\mathbf{u}_k$, the deterministic PWE-CMT-RT framework provides
the received power
$P_{r,k}^{\mathrm{RT}}(\mathbf{x},\mathbf{u}_k)$ according to
\eqref{eq:rt_received_power}. Using the complex PA excitation
$q_m$ defined in \eqref{eq:pa_complex_excitation}, the received
signal at user $k$ is
\begin{equation}
\begin{aligned}
    y_k
    &=
    \left(
        \sum_{m=1}^{M}
        q_m h_{k,m}^{\mathrm{RT}}
    \right)
    \sum_{i=1}^{K}
        \sqrt{\mu_i}\,s_i
    +
    n_k,
\end{aligned}
\label{eq:multiuser_received_signal}
\end{equation}
where $\mu_i\geq 0$ is the power-allocation coefficient of user
$i$, $\sum_{i=1}^{K}\mu_i=1$, and the transmitted symbols satisfy
$\mathbb{E}\{|s_i|^2\}=1$.

Each user decodes its intended signal while treating the signals
intended for the other users as interference. Accordingly, the
SINR of user $k$ is
\begin{equation}
    \gamma_k(\mathbf{x},\boldsymbol{\mu})
    =
    \frac{
        \mu_k
        P_{r,k}^{\mathrm{RT}}(\mathbf{x},\mathbf{u}_k)
    }{
        \displaystyle
        \sum_{\substack{i=1\\i\neq k}}^{K}
        \mu_i
        P_{r,k}^{\mathrm{RT}}(\mathbf{x},\mathbf{u}_k)
        +
        \sigma_k^2
    },
    \label{eq:multiuser_sinr}
\end{equation}
where $\sigma_k^2$ denotes the receiver noise power. Since
$\sum_{i=1}^{K}\mu_i=1$, the SINR can be equivalently expressed as
\begin{equation}
    \gamma_k(\mathbf{x},\boldsymbol{\mu})
    =
    \frac{
        \mu_k
        P_{r,k}^{\mathrm{RT}}(\mathbf{x},\mathbf{u}_k)
    }{
        (1-\mu_k)
        P_{r,k}^{\mathrm{RT}}(\mathbf{x},\mathbf{u}_k)
        +
        \sigma_k^2
    }.
    \label{eq:multiuser_sinr_simplified}
\end{equation}

The achievable spectral efficiency of user $k$ is
\begin{equation}
    R_k(\mathbf{x},\boldsymbol{\mu})
    =
    \log_2
    \left(
        1+\gamma_k(\mathbf{x},\boldsymbol{\mu})
    \right),
    \label{eq:multiuser_user_se}
\end{equation}
measured in bit/s/Hz. The system sum spectral efficiency is
\begin{equation}
    R_{\mathrm{sum}}(\mathbf{x},\boldsymbol{\mu})
    =
    \sum_{k=1}^{K}
    R_k(\mathbf{x},\boldsymbol{\mu}).
    \label{eq:multiuser_sum_se}
\end{equation}

\subsubsection{Joint Optimization Problem}

Maximizing \eqref{eq:multiuser_sum_se} without individual
service requirements may result in a degenerate allocation. In
this case, most or all available power can be assigned to the
user with the most favorable channel. To guarantee a minimum
service level, a common spectral-efficiency requirement
$R_{\min}$ is imposed. Each user must therefore satisfy
$R_k(\mathbf{x},\boldsymbol{\mu})\geq R_{\min}$.

The joint PA-position and user power-allocation problem is
formulated as
\begin{subequations}
\label{prob:multiuser_joint_optimization}
\begin{align}
    \underset{\mathbf{x},\boldsymbol{\mu}}{\operatorname{maximize}}
    \quad&
    R_{\mathrm{sum}}(\mathbf{x},\boldsymbol{\mu})
    \label{prob:multiuser_objective}
    \\
    \operatorname{subject\ to}\quad
    &R_k(\mathbf{x},\boldsymbol{\mu})
    \geq R_{\min},
    \quad \forall k,
    \label{prob:multiuser_qos}
    \\
    &0\leq x_1<x_2\leq L_{\mathrm{wg}},
    \label{prob:multiuser_position}
    \\
    &x_2-x_1\geq d_{\min},
    \label{prob:multiuser_spacing}
    \\
    &\mu_k\geq0,
    \quad \forall k,
    \label{prob:multiuser_power_nonnegative}
    \\
    &\sum_{k=1}^{K}\mu_k=1.
    \label{prob:multiuser_power_sum}
\end{align}
\end{subequations}

Problem~\eqref{prob:multiuser_joint_optimization} is challenging because
$P_{r,k}^{\mathrm{RT}}$ has no closed-form dependence on the PA positions,
which jointly govern sequential waveguide-power extraction and multipath
propagation, while coupling with $\boldsymbol{\mu}$ through the SINR and
QoS constraints.

\subsubsection{Physics-Assisted Electromagnetic Surrogates}

Exhaustive evaluation using the full electromagnetic models is
computationally prohibitive. We therefore precompute the waveguide response
and single-source wireless channels separately and cascade their surrogates
for each candidate PA configuration.

For the waveguide stage, the PWE-CMT solution is exported as
position-dependent power and phase lookup curves. Let
$a_{\mathrm{base}}(x)$ denote the complex guided-mode coefficient
extracted from the full PWE field at longitudinal position $x$,
and let $a_{\mathrm{in}}$ denote its input reference. The baseline
power ratio is
\begin{equation}
    b(x)
    =
    \left|
        \frac{
            a_{\mathrm{base}}(x)
        }{
            a_{\mathrm{in}}
        }
    \right|^2.
    \label{eq:pwe_baseline_ratio}
\end{equation}
For field-retention coefficients $\alpha_1$ and $\alpha_2$, the
normalized PA radiation coefficients are
\begin{equation}
    \chi_1(\mathbf{x})
    =
    b(x_1)
    \left(
        1-\alpha_1^2
    \right)
    \label{eq:pwe_chi1}
\end{equation}
and
\begin{equation}
    \chi_2(\mathbf{x})
    =
    b(x_2)
    \alpha_1^2
    \left(
        1-\alpha_2^2
    \right).
    \label{eq:pwe_chi2}
\end{equation}
The corresponding radiation powers are
\begin{equation}
    P_{\mathrm{rad},m}
    =
    P_{\max}\chi_m(\mathbf{x}).
    \label{eq:surrogate_pa_radiation_power}
\end{equation}
Since the positive real coefficient $\alpha_m$ introduces no additional
phase, the radiation phase is
\begin{equation}
    \theta_{\mathrm{rad},m}
    =
    \arg\left(
        -j a_{\mathrm{base}}(x_m)
    \right).
    \label{eq:surrogate_pa_radiation_phase}
\end{equation}
The resulting lookup therefore preserves position-dependent radiation
power, sequential guided-power depletion, and accumulated phase.

For the wireless stage, a shared physics-assisted RT surrogate maps
$\mathbf{c}_{k,m}
=[x_m,x_k^{\mathrm u},y_k^{\mathrm u}]^{\mathrm T}$
to the single-source channel
$\widehat h_{k,m}^{\mathrm{RT}}$. Each PA-user pair is evaluated
independently by the same surrogate, while multi-PA coherent
combining is performed only after the single-source channels have
been predicted.

Each candidate topology $\ell$ specifies an ordered sequence of reflecting
surfaces, including the line-of-sight case. The image-source module returns
its path length $d_\ell(\mathbf{c}_{k,m})$ and visibility indicator
$A_\ell(\mathbf{c}_{k,m})$, which equals one only when all reflection points
lie on their prescribed finite surfaces and every path segment is unblocked.

For each feasible topology, a shared neural decoder predicts a
visibility logit $z_\ell$ and a phase-stripped complex path weight
$\widehat b_\ell$. Its effective contribution is controlled by
\begin{equation}
    \rho_\ell
    =
    A_\ell(\mathbf{c}_{k,m})
    \sigma(z_\ell),
    \label{eq:topology_gate}
\end{equation}
where the analytical factor enforces geometric feasibility and
the learned factor accounts for residual propagation effects.
The carrier phase is restored from the analytical path length,
and the single-source channel is reconstructed as
\begin{equation}
    \widehat h_{k,m}^{\mathrm{RT}}
    =
    \sum_{\ell\in\mathcal{S}_{k,m}}
    \rho_\ell
    \widehat b_\ell
    \exp\left(
        -j
        \frac{2\pi f_{\mathrm{c}}}{c_0}
        d_\ell(\mathbf{c}_{k,m})
    \right),
    \label{eq:rt_channel_reconstruction}
\end{equation}
where $\mathcal{S}_{k,m}$ is the retained topology set and $c_0$
is the speed of light. This physics-assisted reconstruction
avoids directly learning the rapidly varying carrier phase.

The surrogate is trained using the weighted multi-task loss
\begin{equation}
\begin{aligned}
    \mathcal{L}_{\mathrm{RT}}
    ={}&
    \lambda_{\mathrm{vis}}\mathcal{L}_{\mathrm{vis}}
    +
    \lambda_{\mathrm{rank}}\mathcal{L}_{\mathrm{rank}}
    +
    \lambda_b\mathcal{L}_b
    +
    \lambda_h\mathcal{L}_h.
\end{aligned}
\label{eq:rt_surrogate_loss}
\end{equation}
The four loss terms supervise path visibility, topology ranking,
phase-stripped path weights, and end-to-end complex-channel reconstruction,
respectively. The implementation and accuracy are reported in the
Numerical Results section.

For a candidate PA configuration, the predicted single-source
channels are combined with the precomputed PWE-CMT radiation
powers and phases as
\begin{equation}
\begin{aligned}
    \widehat P_{r,k}^{\mathrm{RT}}
    &=
    P_{\max}
    \left|
        \sum_{m=1}^{M}
        \sqrt{\chi_m(\mathbf{x})}\,
        \exp\left(
            j\theta_{\mathrm{rad},m}
        \right)
        \widehat h_{k,m}^{\mathrm{RT}}
    \right|^2.
\end{aligned}
\label{eq:surrogate_received_power}
\end{equation}
Thus, both the PWE-CMT source phases and the RT multipath phases
are retained in the coherent combination. The corresponding
SINRs and SEs follow from
\eqref{eq:multiuser_sinr}--\eqref{eq:multiuser_sum_se}.

\subsubsection{Supervised Coarse-to-Fine Optimization}

The feasible PA-placement space is discretized into a fine grid
and partitioned into coarse regions. Predicting a single
fine-grid placement is undesirable because several physically
distinct configurations may achieve nearly identical objective
values. We therefore train the placement predictor to rank a set
of near-optimal coarse regions.

Teacher labels are generated offline from random feasible user
configurations. For each user set $\mathcal U$, all fine-grid PA
configurations are evaluated using the cascaded surrogate, with
the QoS-constrained power-allocation subproblem solved for every
candidate. Let $J_j(\mathcal U)$ denote the resulting sum SE of
fine-grid candidate $j$, and let
$J^\star(\mathcal U)$ be the largest feasible value. The
$\epsilon$-optimal fine-grid set is
\begin{equation}
    \mathcal E(\mathcal U)
    =
    \left\{
        j:
        J^\star(\mathcal U)-J_j(\mathcal U)
        \leq\epsilon
    \right\}.
    \label{eq:epsilon_optimal_set}
\end{equation}
These candidates are mapped to their corresponding coarse
regions. If several candidates map to the same region, the
largest associated sum SE is used as its teacher objective,
denoted by $\overline J_r(\mathcal U)$ for region $r$.

To ensure invariance to user ordering, a shared encoder maps each
user coordinate to a latent feature, followed by mean aggregation:
\begin{equation}
    \mathbf z_{\mathrm g}
    =
    \frac{1}{K}
    \sum_{k=1}^{K}
    f_{\mathrm{enc}}(\mathbf u_k).
    \label{eq:placement_global_feature}
\end{equation}
The predictor first estimates the coarse index of $x_1$ and then
predicts that of $x_2$ conditioned on $\mathbf z_{\mathrm g}$ and
the first-position index. Invalid or spacing-violating pairs are
masked before normalization.

Let $\mathcal C_\epsilon(\mathcal U)$ denote the coarse regions
induced by $\mathcal E(\mathcal U)$. Their objective-dependent
weights are
\begin{equation}
    w_r
    =
    \frac{
        \exp\left(
            \left[
                \overline J_r-J^\star
            \right]/\tau
        \right)
    }{
        \displaystyle
        \sum_{i\in\mathcal C_\epsilon(\mathcal U)}
        \exp\left(
            \left[
                \overline J_i-J^\star
            \right]/\tau
        \right)
    },
    \label{eq:teacher_region_weight}
\end{equation}
where $\tau$ is a temperature parameter. The placement predictor
is trained using
\begin{equation}
    \mathcal L_{\mathrm{place}}
    =
    -\frac{1}{N_{\mathrm B}}
    \sum_{n=1}^{N_{\mathrm B}}
    \sum_{r\in\mathcal C_\epsilon(\mathcal U^{(n)})}
    w_r^{(n)}
    \log
    q_{\boldsymbol\psi}
    \left(
        r\mid\mathcal U^{(n)}
    \right),
    \label{eq:placement_predictor_loss}
\end{equation}
where $q_{\boldsymbol\psi}(r\mid\mathcal U)$ is the predicted
probability of coarse region $r$ and $N_{\mathrm B}$ is the
mini-batch size. This set-valued objective avoids over-penalizing
distinct placement regions with nearly identical sum SEs.
Training settings and grid resolutions are provided in the
Numerical Results section.

During inference, the highest-ranked coarse regions are expanded into
feasible fine-grid candidates, reevaluated by the cascaded surrogates, and
ranked after optimizing their power allocations. A shortlist is then
validated by RT simulation to mitigate ranking errors near coherent
cancellation. Power allocation is reoptimized using the reference
channels, and the feasible configuration with the largest refined sum SE
is selected. The complete workflow is shown in Fig.~\ref{fig:suanfa}.

% =========================================================
% Section IV
% =========================================================
\section{Numerical Results}
\label{sec:numerical_results}

\begin{figure}[htbp]
    \centering
    \includegraphics[width=0.75\linewidth]{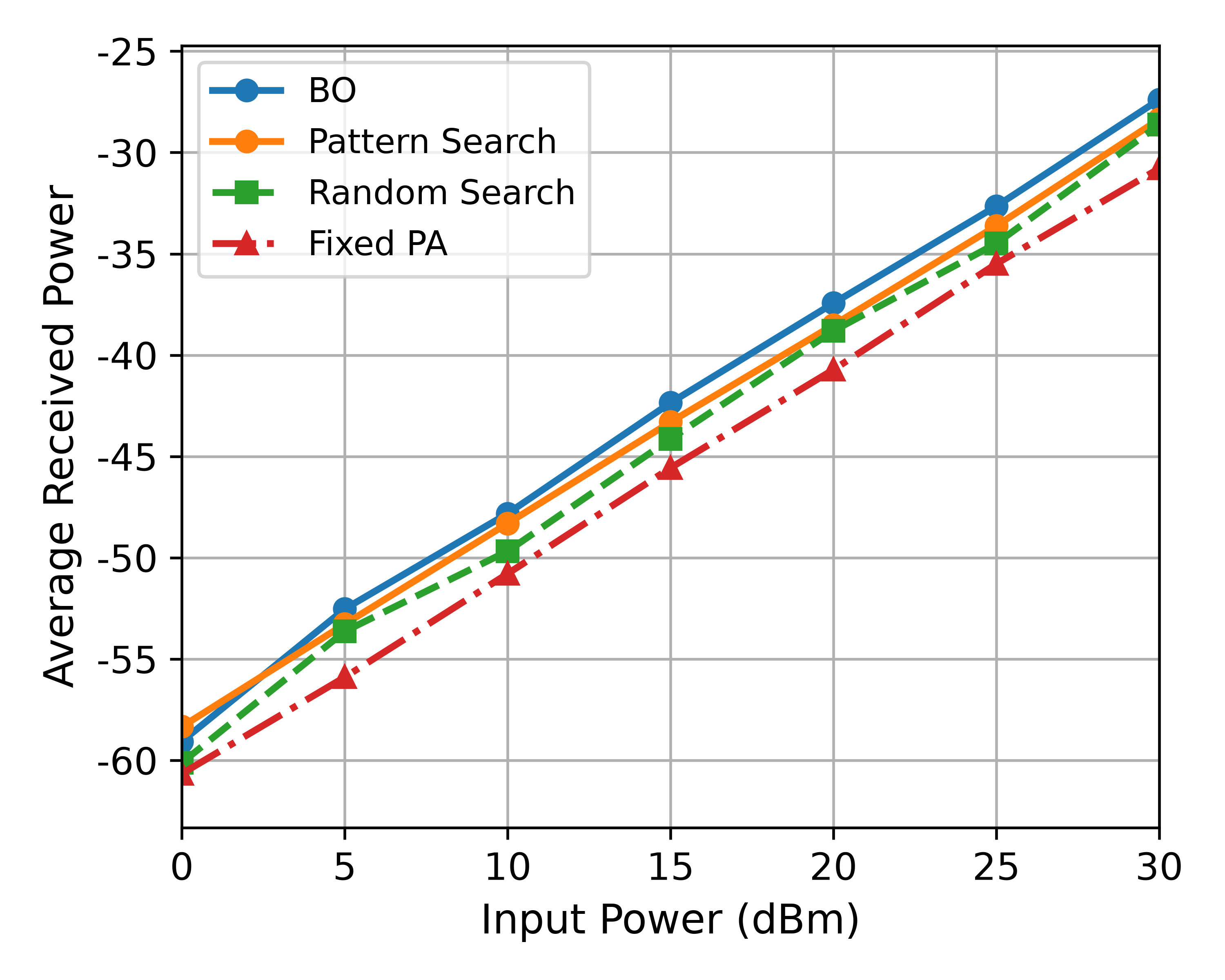}
    \caption{Average received power of Bayesian optimization and baseline methods across transmit power levels (0-30~dBm).}
    \label{fig:bo_baseline_path_loss}
\end{figure}

\subsection{Single-User Scenario Optimization}

\subsubsection{Simulation Environment and BO Settings}

The single-user evaluation is conducted in the site-specific
indoor environment described in
Section~\ref{sec:physics_modeling}. A 19.5 m dielectric waveguide
is deployed in the environment, and the carrier frequency is
28~GHz. The PWE-CMT model determines the PA radiation power
and phase, while a site-specific RT solver evaluates indoor propagation. User locations are sampled from the valid
receiver area, excluding locations occupied by obstacles.

For each user location $\mathbf{u}$, the PA position is optimized
over the waveguide according to
\begin{equation}
    x_p^\star(\mathbf{u})
    =
    \underset{0\leq x_p\leq L_{\mathrm{wg}}}
    {\operatorname{arg\,max}}\;
    P_{r,\mathrm{dBm}}^{\mathrm{RT}}
    \left(
        x_p,\mathbf{u}
    \right).
    \label{eq:single_user_bo_objective}
\end{equation}
Each objective evaluation invokes the deterministic PWE-CMT-RT
framework. BO is employed to reduce the number of these
computationally expensive evaluations. It uses five initial
samples followed by 25 iterations, corresponding to a total
budget of 30 evaluations. The complete settings are listed in
Table~\ref{tab:BO_hyperparameters}.

\begin{figure*}[t]
    \centering
    \includegraphics[width=\textwidth]{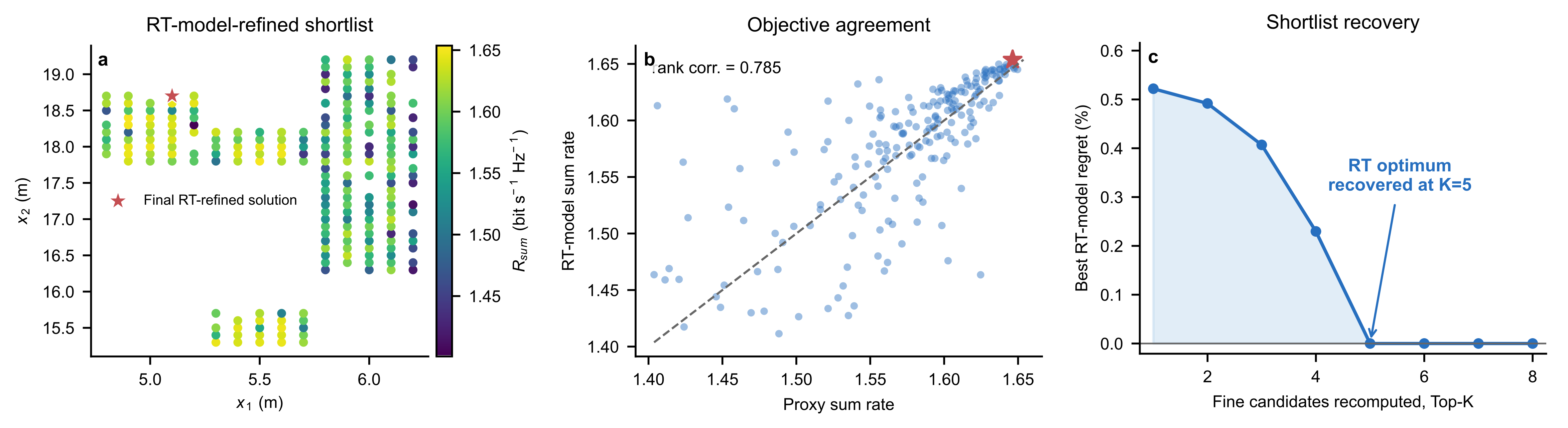}
    \caption{Validation of the QoS-constrained coarse-to-fine
optimization. (a) RT-refined sum SE over the fine-grid PA pairs
expanded from the Top-10 coarse shortlist. Each point represents
a PA pair $(x_1,x_2)$, and its color indicates the RT-refined sum
SE, from low (dark purple) to high (yellow); the red star marks
the final PA placement. (b) Agreement between the surrogate and
reference RT objectives over jointly feasible candidates.
(c) Regret after recomputing the Top-$K_{\mathrm f}$
surrogate-ranked fine-grid candidates using reference RT.}
    \label{fig:optimization_validation}
\end{figure*}

\begin{table}[htbp]
\centering
\footnotesize
\caption{Bayesian optimization settings.}
\label{tab:BO_hyperparameters}
\begin{tabularx}{\columnwidth}{
    >{\centering\arraybackslash}X
    >{\centering\arraybackslash}X
    >{\centering\arraybackslash}X}
\toprule
Parameter & Meaning & Value \\
\midrule
\texttt{init\_points} & Initial samples & 5 \\
\texttt{n\_iter} & BO iterations & 25 \\
\texttt{random\_state} & Random seed & 42 \\
\texttt{acquisition} & Acquisition function & EI \\
\texttt{kernel} & GP kernel & Matérn 5/2 \\
$\alpha$ & GP noise variance & 0 \\
\shortstack{\texttt{n\_restarts}\\
\texttt{\_optimizer}}
& Kernel-fitting restarts & 5 \\
\bottomrule
\end{tabularx}
\end{table}

\subsubsection{Single-User Optimization Results}

BO is compared with pattern search, random search, and a fixed-PA
strategy. For each transmit-power level, 1,000 user locations are
independently sampled. All methods are evaluated using the same
user locations and the same objective-evaluation budget. Because
the optimal PA position varies with the user location, the
Monte Carlo experiment assesses the average placement performance
rather than identifying one universal PA position.

Fig.~\ref{fig:bo_baseline_path_loss} compares the average received
power achieved by the four methods. BO achieves the highest value
at six of the seven transmit-power levels, while pattern search
performs slightly better at 0~dBm. Averaged over all power levels,
BO outperforms pattern search, random search, and the fixed-PA
strategy by 0.641, 1.465, and 2.952~dB, respectively. Its gain
over the fixed-PA strategy ranges from 1.594 to 3.396~dB. These
results indicate that BO provides the best average PA-placement
performance for randomly distributed users under the prescribed
evaluation budget.

\subsection{Multi-User Scenario Optimization}

\subsubsection{Simulation Setup}
\label{subsec:simulation_setup}

We considered an indoor environment with dimensions
$19.5\times7.5\times3.4~\mathrm{m}^{3}$. The waveguide was installed along
the wall at $y=0.01~\mathrm{m}$ and $z=3.0~\mathrm{m}$, whereas all receivers
were placed at $z=1.3~\mathrm{m}$. Receiver locations overlapping the indoor
obstacles were excluded. The carrier frequency was $28~\mathrm{GHz}$, and the
transmitters and receivers employed single isotropic vertically polarized
antennas. The RT calculation retained line-of-sight propagation, specular
reflection, and diffraction up to a maximum interaction depth of three. The
principal simulation and optimization parameters are summarized in
Table~\ref{tab:simulation_parameters}.

\begin{table}[t]
    \centering
    \caption{Principal simulation and optimization parameters.}
    \label{tab:simulation_parameters}
    \setlength{\tabcolsep}{4.0pt}
    \begin{tabular}{@{}ll@{}}
        \toprule
        Parameter & Value \\
        \midrule
        Carrier frequency, $f_{\mathrm c}$ & $28~\mathrm{GHz}$ \\
        Relative permittivity & $5.31$ \\
        Conductivity & $0.8967~\mathrm{S/m}$ \\
        Maximum RT depth & $3$ \\
        Samples per source & $4\times10^{6}$ \\
        Waveguide length, $L_{\mathrm{wg}}$ & $19.5~\mathrm{m}$ \\
        PA field-retention coefficients & $\alpha_1=\alpha_2=0.9$ \\
        Minimum PA separation, $d_{\min}$ & $0.5~\mathrm{m}$ \\
        Fine PA-position resolution & $0.1~\mathrm{m}$ \\
        Waveguide input power, $P_{\max}$ & $1~\mathrm{W}$ \\
        Bandwidth & $1~\mathrm{GHz}$ \\
        Receiver noise figure & $7~\mathrm{dB}$ \\
        Minimum user SE, $R_{\min}$ & $0.35~\mathrm{bit/s/Hz}$ \\
        Number of legal fine-grid PA pairs & $18{,}336$ \\
        Number of coarse PA regions & $780$ \\
        Number of teacher configurations & $5{,}000$ \\
        \bottomrule
    \end{tabular}
\end{table}

The receiver noise power was calculated using a thermal-noise density of
$-174~\mathrm{dBm/Hz}$, a bandwidth of $1~\mathrm{GHz}$, and a receiver noise
figure of $7~\mathrm{dB}$. The resulting value was used consistently during
teacher-label generation, surrogate-based candidate evaluation, and RT-model
refinement.

The in-waveguide propagation was simulated using the
split-step-Fourier implementation of the PWE. The transverse computational
window covered $0.2\times0.2~\mathrm{m}^{2}$ and was discretized
using $\Delta y=\Delta z=\lambda/4$, where $\lambda$ is the
free-space wavelength. An impedance boundary condition was
applied along the transverse boundary to account for field
confinement and finite boundary loss.

The path-level RT dataset contained $50{,}000$ Tx-Rx pairs and approximately
$2.30\times10^{6}$ propagation paths. It included global, local, and boundary
samples. Each propagation path was represented by its complex coefficient,
delay, path length, interaction type, and reflection geometry. The
physics-assisted shared-topology model was evaluated on an independent test set
of $5{,}000$ Tx-Rx links.

The placement-prediction dataset contained $5{,}000$ four-user configurations,
of which $4{,}058$, $451$, and $491$ were assigned to the training,
validation, and test sets, respectively. For each teacher configuration, all
$18{,}336$ legal fine-grid PA pairs were evaluated using the cascaded PWE-RT
surrogate. The QoS-constrained power-allocation subproblem was solved for every
fixed PA pair. All teacher configurations admitted at least one feasible
solution, and the mean fraction of feasible PA pairs was $0.563$.

\subsubsection{Fidelity of the Physics-Assisted RT Surrogate}
\label{subsec:rt_surrogate_results}

The held-out evaluation shows that the single-source complex-channel
surrogate achieves an NMSE of $-19.1436\,\mathrm{dB}$ and a complex
correlation coefficient of $0.9940$. Across all test links, the median and
95th-percentile absolute phase errors were $0.05053$ and $0.43844~\mathrm{rad}$,
respectively. For strong channels, these values decreased to $0.04649$ and
$0.30270~\mathrm{rad}$.

After model loading, a neural forward evaluation required approximately
$2~\mathrm{ms}$, whereas a single reference RT link calculation required
approximately $2.5~\mathrm{s}$ under the adopted configuration. The complete
single-link neural routine, including data preparation and model-handling
overheads, required approximately $77~\mathrm{ms}$. The surrogate therefore
provides the computational efficiency required to examine a large PA-placement
space. However, small channel-phase errors can noticeably affect the
predicted received power when different PA contributions nearly
cancel coherently. The neural surrogate is therefore used to
screen the full placement space, whereas the shortlisted
candidates are reevaluated using the reference RT model before
the final design is selected.

\begin{figure}[h]
    \centering
    \includegraphics[width=\columnwidth]{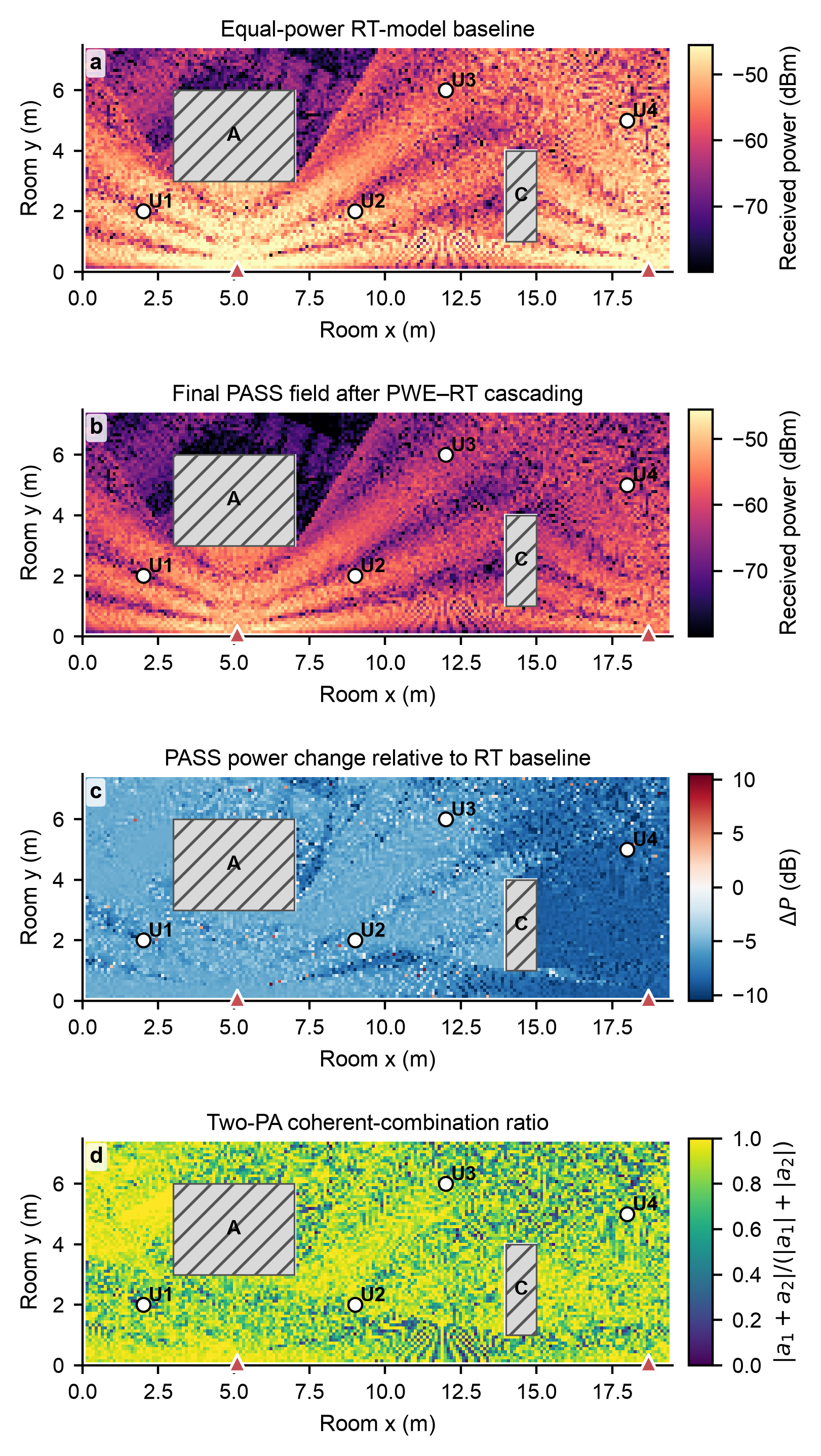}
    \caption{Indoor field distributions at the final RT-refined
    PA placement. (a) Equal-power RT baseline. (b) PASS received
    power after ordered PWE weighting and coherent PWE-RT
    cascading. (c) Received-power difference between the complete
    PASS model and the equal-power baseline. (d) Coherent-
    combination ratio of the two PA contributions. Hatched
    regions denote inaccessible obstacles, triangles denote the
    PA positions, and circles denote the users.}
    \label{fig:optimized_field}
\end{figure}

\subsubsection{QoS-Constrained Multi-User Optimization}
\label{subsec:optimization_results}

We evaluate the proposed method for four users located at
$(2.0,2.0)$, $(9.0,2.0)$, $(12.0,6.0)$, and
$(18.0,5.0)~\mathrm{m}$. Each user is required to satisfy
$R_k\geq R_{\min}=0.35~\mathrm{bit/s/Hz}$.

Given the user configuration, the predictor selects the Top-$K_{\mathrm c}$
coarse regions. Their legal fine-grid PA pairs are ranked by the cascaded
PWE-RT surrogate after QoS-constrained power optimization, and the
highest-ranked candidates are refined using reference RT.

\begin{figure}[h]
    \centering
    \includegraphics[width=0.45\textwidth]{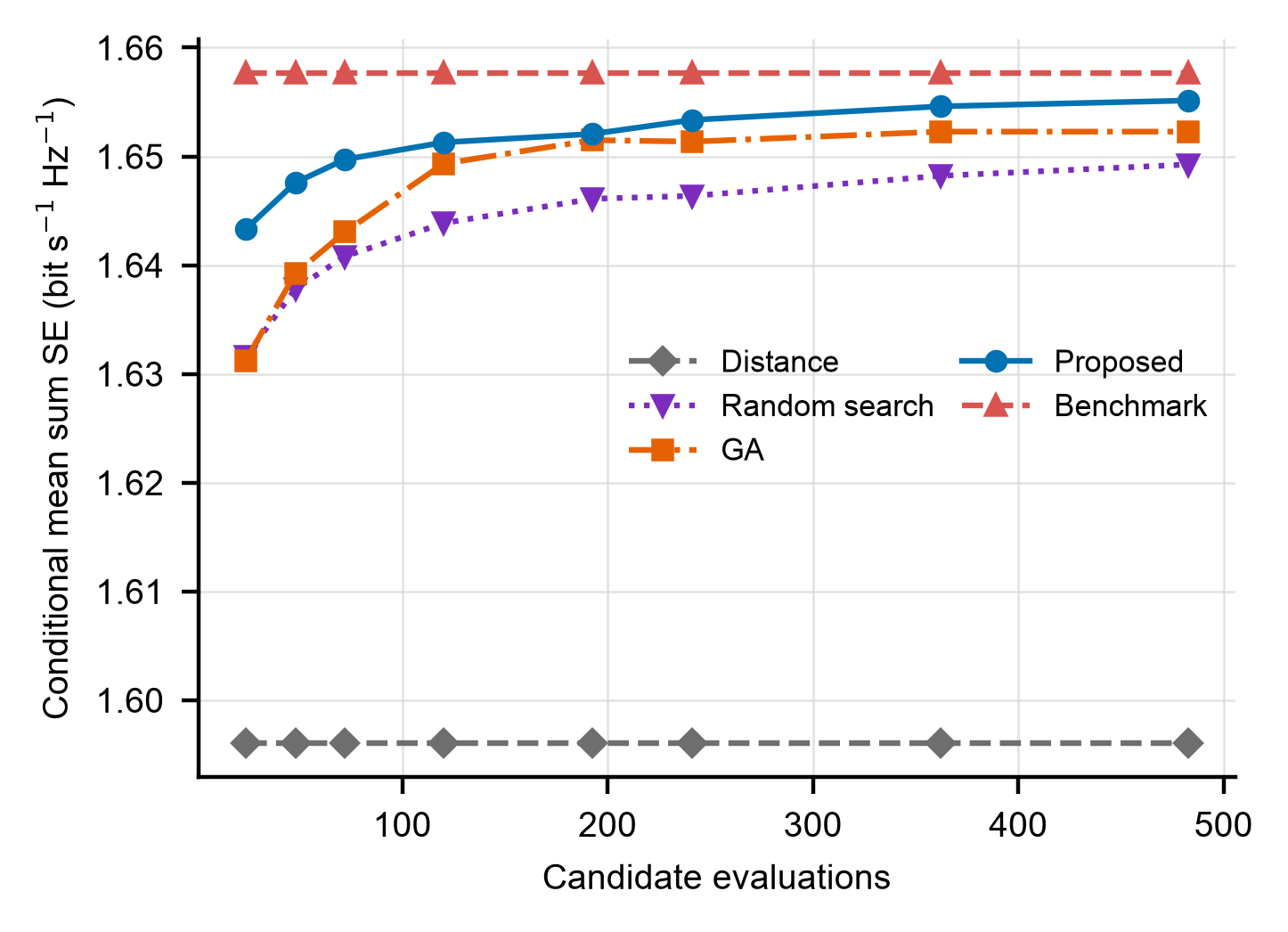}
    \caption{Conditional mean sum spectral efficiency versus the candidate-evaluation budget over 491 held-out user configurations.}
    \label{fig:compare}
\end{figure}

The coarse-region predictor was evaluated on 491 held-out
four-user configurations. Table~\ref{tab:shortlist_audit} reports
its feasible recall and 95th-percentile sum-SE regret for different
values of $K_{\mathrm c}$. Feasible recall is the fraction of test
configurations for which the expanded shortlist contains at least
one QoS-feasible fine-grid solution. Regret is measured relative
to exhaustive fine-grid evaluation using the cascaded surrogate.
Increasing $K_{\mathrm c}$ improves both feasibility recovery and
decision quality by retaining more candidate regions.

\begin{table}[t]
    \centering
    \caption{Held-out decision quality of the learned
    coarse-region predictor.}
    \label{tab:shortlist_audit}
    \setlength{\tabcolsep}{4.0pt}
    \begin{tabular}{@{}lcc@{}}
        \toprule
        Coarse shortlist
        & 95th-percentile regret
        & Feasible recall \\
        & (bit/s/Hz) & \\
        \midrule
        Top-1  & $0.08669$ & $0.97556$ \\
        Top-5  & $0.03253$ & $0.99389$ \\
        Top-10 & $0.02070$ & $0.99796$ \\
        Top-20 & $0.01453$ & $0.99796$ \\
        \bottomrule
    \end{tabular}
\end{table}

For the representative user configuration, the Top-10 coarse
shortlist expands to 250 legal fine-grid PA pairs. Each colored
point in Fig.~\ref{fig:optimization_validation}(a) represents one
candidate pair, with its horizontal and vertical coordinates
giving $x_1$ and $x_2$, respectively. The color indicates the sum
SE obtained after RT-model refinement: lighter yellow colors
denote higher values, whereas darker purple colors denote lower
values. The final RT-refined PA placement is
\begin{equation}
    (x_1^\star,x_2^\star)
    =
    (5.1,18.7)~\mathrm{m}.
    \label{eq:final_refined_pa_positions}
\end{equation}

Figure~\ref{fig:optimization_validation}(b) compares the surrogate and
reference RT objectives for jointly feasible candidates, yielding a rank
correlation of $0.7851$. After RT recomputation, the best
surrogate-ranked candidate has a relative regret of $0.522\%$, while
$8.0\%$ of the shortlisted candidates are incorrectly classified as
QoS-feasible, motivating the final RT refinement.

In Fig.~\ref{fig:optimization_validation}(c), Top-$K_{\mathrm f}$
refinement means that only the $K_{\mathrm f}$ highest-ranked
fine-grid candidates from the surrogate are recomputed using the
reference RT model. The final RT-refined solution is recovered
when $K_{\mathrm f}=5$, showing that only a small fine-grid
shortlist is required. 

At the final solution, the optimized power-allocation vector is
\[
\boldsymbol{\mu}^{\star}
=
[0.34471,\;0.21750,\;0.22018,\;0.21761]^{\mathrm T},
\]
and the resulting user SEs are
\[
\mathbf{R}^{\star}
=
[0.60369,\;0.35000,\;0.35000,\;0.35000]^{\mathrm T}
~\mathrm{bit/s/Hz}.
\]
The corresponding sum SE is
$1.65369~\mathrm{bit/s/Hz}$. Three users operate at the QoS
boundary, while the remaining power is assigned to the user that
provides the largest increase in sum SE.

Fig.~\ref{fig:optimized_field} shows the spatial received-field
characteristics at the final RT-refined PA placement. All maps
are evaluated over the receiver plane. Panel (a) presents an
equal-power RT baseline that assigns the same radiation power to
both PAs without applying the position-dependent PWE-CMT
weighting. Panel (b) shows the received power predicted by the
complete PASS model, which incorporates the ordered PA radiation
powers, PA source phases, and coherent multipath combination.

Panel (c) shows the received-power difference between the complete
PASS model and the equal-power RT baseline. Positive values
indicate that the PWE-CMT-weighted PASS model produces higher
received power than the baseline, while negative values indicate
a reduction. 

Panel (d) quantifies the degree of coherent combination between
the two PA contributions.
Let
$a_m(\mathbf{r})=\sqrt{P_{\mathrm{rad},m}}
e^{j\theta_{\mathrm{rad},m}}
h_m^{\mathrm{RT}}(\mathbf{r})$
denote the complex contribution of the $m$-th PA at
$\mathbf{r}$. The coherent-combination ratio is defined as
\begin{equation}
    \eta_{\mathrm{coh}}(\mathbf{r})
    =
    \frac{
        \left|
            a_1(\mathbf{r})+a_2(\mathbf{r})
        \right|
    }{
        \left|a_1(\mathbf{r})\right|
        +
        \left|a_2(\mathbf{r})\right|
    }.
    \label{eq:coherent_combination_ratio}
\end{equation}
The ratio satisfies
$0\leq\eta_{\mathrm{coh}}(\mathbf{r})\leq1$. Values approaching
one indicate that the two PA contributions combine
constructively, whereas values approaching zero indicate strong
coherent cancellation. Accordingly, brighter yellow colors in
panel (d) represent stronger constructive combination, while
darker purple colors identify cancellation-sensitive regions.

\begin{figure*}[t]
    \centering
    \includegraphics[width=0.85\textwidth]
    {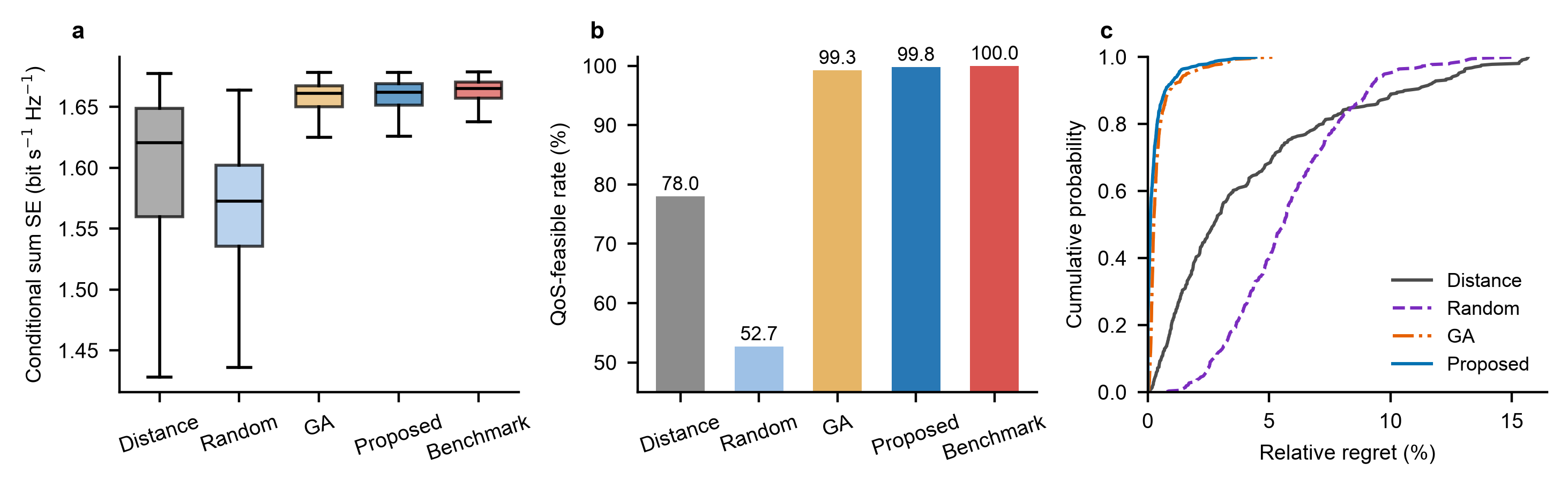}
    \caption{Comparison of the PA-placement search algorithms over \(491\) held-out user configurations. 
    (a) Conditional sum spectral efficiency over QoS-feasible trials.
    (b) QoS-feasible rate.
    (c) Empirical cumulative distribution of the relative regret over QoS-feasible trials.}
    \label{fig:algorithm_comparison}
\end{figure*}

The received-power distribution is spatially nonmonotonic because
it is jointly determined by indoor multipath propagation, ordered
waveguide power extraction, and coherent combination of the PA
contributions. Consequently, locations with similar propagation
distances may experience substantially different received powers.

The proposed coarse-to-fine pipeline preserves this phase-resolved
electromagnetic behavior while reducing online computation. Only
the fine-grid candidates within the predicted coarse regions are
evaluated using the physical models, thereby avoiding exhaustive
online evaluation of the complete PA-placement space.

\subsubsection{Comparison with Placement-Search Baselines}
\label{subsubsec:algorithm_comparison}

We compare the proposed strategy with random placement, a distance-based
heuristic, a genetic algorithm (GA), and exhaustive search. All methods
use the same deterministic PWE-RT channels and QoS-constrained
power-allocation solver, thereby isolating the effect of placement
search. Random placement uniformly samples one legal configuration,
whereas the distance-based heuristic minimizes each user's distance to
its nearest PA. The GA and the proposed method use the same
candidate-evaluation budget, while exhaustive search evaluates all
$18{,}336$ legal configurations on the $0.1$ m grid.

The comparison was conducted over \(491\) held-out user configurations.  As shown in Fig. \ref{fig:compare}, the proposed method consistently achieves a higher conditional sum SE than the GA and random search under matched candidate budgets. As shown in Fig.~\ref{fig:algorithm_comparison}(a), the proposed method achieved a mean conditional sum spectral efficiency of \(1.6534~\mathrm{bit\,s^{-1}\,Hz^{-1}}\), close to the benchmark value of \(1.6577~\mathrm{bit\,s^{-1}\,Hz^{-1}}\). The corresponding result of the GA was \(1.6516~\mathrm{bit\,s^{-1}\,Hz^{-1}}\). Here, the conditional sum spectral efficiency was calculated only over QoS-feasible trials; infeasible outcomes were accounted for separately through the feasibility rate.

Fig.~\ref{fig:algorithm_comparison}(b) shows that the proposed method attained a QoS-feasible rate of \(99.8\%\), compared with \(99.3\%\) for the GA, \(78.0\%\) for the distance-based method, and \(52.7\%\) for random placement. The exhaustive-search benchmark was feasible for all evaluated user configurations. These results indicate that purely geometric proximity is insufficient for selecting PA positions because it does not capture the position-dependent waveguide power extraction, multipath phase, and coherent superposition of the PA signals.

To quantify the optimality loss, the relative regret was defined as
\begin{equation}
\mathcal{G}
=
\frac{
R_{\mathrm{sum}}^{\mathrm{bench}}
-
R_{\mathrm{sum}}^{\mathrm{alg}}
}{
R_{\mathrm{sum}}^{\mathrm{bench}}
}
\times 100\%,
\end{equation}
where \(R_{\mathrm{sum}}^{\mathrm{bench}}\) denotes the maximum sum spectral efficiency obtained by exhaustive search over the discretized candidate set, and \(R_{\mathrm{sum}}^{\mathrm{alg}}\) denotes that obtained by the evaluated algorithm. The regret statistics were calculated over QoS-feasible trials. As shown in Fig.~\ref{fig:algorithm_comparison}(c), the proposed method produced the most concentrated regret distribution among the non-exhaustive methods. Its mean and \(95\)th-percentile relative regrets were \(0.293\%\) and \(1.261\%\), respectively, compared with \(0.435\%\) and \(1.556\%\) for the GA. By contrast, the distance-based and random-placement methods yielded mean regrets of \(4.169\%\) and \(5.378\%\), respectively.

The proposed method and the GA evaluated approximately \(241\) candidates per user configuration, corresponding to only \(1.32\%\) of the \(18{,}336\) candidate search space. Despite their similar spectral-efficiency performance, the proposed method incurred substantially lower online search overhead. Excluding the common physical-channel and objective-function evaluations, its mean search time was \(0.628~\mathrm{ms}\), whereas the GA required \(22.59~\mathrm{ms}\). The proposed method therefore reduced the placement-search latency by approximately \(36\) times under the matched candidate budget. This advantage arises because the trained predictor directly identifies promising coarse regions in a single forward pass, while the GA requires iterative population evolution. Overall, the proposed strategy approached the discretized exhaustive-search benchmark while maintaining a high QoS-feasibility rate and substantially lower online search complexity.

\section{Conclusion}

This paper presented a physics-driven deterministic framework for
end-to-end modeling and optimization of indoor PASS. The proposed
PWE-CMT-RT framework jointly captures in-waveguide field evolution,
sequential PA coupling and power extraction, and site-specific multipath
propagation. Phase-consistent combination of these components enables the
received power and spatial radio map to be evaluated while preserving
interference among propagation paths and PAs. To reduce the cost of
repeated physical simulations, we developed a reusable PWE-CMT field
representation and a physics-assisted RT surrogate, with reference RT
retained for final refinement. Numerical results demonstrated high
complex-channel fidelity and substantially reduced evaluation time. The
framework was further applied to single- and multi-user scenarios for
joint PA-placement and power-allocation design, demonstrating that
phase-resolved electromagnetic modeling can be incorporated into
communication-oriented PASS optimization. Future work will investigate
larger systems, experimental validation, higher-fidelity electromagnetic
models, and PASS architectures with alternative waveguides, feeding
mechanisms, and propagation modes.

\bibliographystyle{IEEEtran}
\bibliography{ref}

% if have a single appendix:
%\appendix[Proof of the Zonklar Equations]
% or
%\appendix  % for no appendix heading
% do not use \section anymore after \appendix, only \section*
% is possibly needed

% use appendices with more than one appendix
% then use \section to start each appendix
% you must declare a \section before using any
% \subsection or using \label (\appendices by itself
% starts a section numbered zero.)
%

% that's all folks
\end{document}